\documentstyle[twocolumn,prl,aps,floats]{revtex}

\begin{document}
\draft
\title{Calculation of the interaction of a neutron spin with an atomic electric
field}
\author{R. Golub$^1$, S.K. Lamoreaux$^2$}
\address{1. Hahn-Meitner Institute, Glienicker Str. 100, D14109 Berlin,\\
Germany\\
2. Los Alamos National Laboratory, Physics Division, P-23,\\
M.S. H803, Los Alamos, NM 87545}

\date{\today}

\maketitle

\tightenlines

\begin{abstract}

The Thomas-Fermi approximation for an atomic wavefunction is used to
calculate the interaction of a neutron spin with the atomic electric field,
either through the motional 
magnetic  ($\vec v\times \vec E$) or possibly electric (due to
the possible existence of a neutron permanent electric dipole moment)
couplings.
\end{abstract}


\input{psfig.sty}

\section{Introduction}

Most recent experiments to search for the neutron electric dipole moment
(EDM) involve the neutron interacting with electric fields created by
laboratory apparatus. However there is also the possibility of using
electric fields produced by atoms in crystals. There are some reasons for
believing this might be advantageous -- atomic fields are large and the
neutrons interact with many atoms in a coherent fashion. However as we will
show the measurable effects are quite small compared to known background
effects due to the motional electric field.

The effect in a crystal experiment can be enhanced if the scatteing
amplitude due to the EDM interaction can be made to interfere with the much
larger scattering amplitude of a nucleus. It will be seen below that the
scattering amplitude due to the EDM interaction is imaginary so it can only
interfere with an imaginary nuclear amplitude. The idea of searching for a
neutron  EDM by measuring the interference between the scattering from an
atomic electric field (due to the EDM interaction) with nuclear scattering
was proposed by Shull; an experiment was performed in 1964 and was based on
scattering from a CdS crystal, because Cd, a strong absorber, has a large
imaginary scattering amplitude.\cite{sch}  In this experiment, the
penetration depth, hence reflectivity, depends on the orientation of the
neutron spin relative to the momentum transfer.\cite{gol,lam}

Recently, a proposal to search for a neutron EDM by scattering in a perfect
Si crystal was put forward.\cite{dom} In this case, the imaginary part of
the nuclear scattering length is very small, and the proposed observable is
a rotation of the neutron spin direction caused by the superpostion of the
spin dependent imaginary amplitude with the real nuclear amplitude.

Because the calculations regarding this effect do not, to our knowledge,
appear in the literature and the calculations regarding the Schull
experiment are lacking in detail, we felt it worthwhile to estimate the size
of a neutron spin rotation due to an EDM interaction, and in addition,
include the analysis of the $\vec{v}\times \vec{E}$ interaction originally
studied by Schwinger \cite{schw} and demonstrated by 
Shull and Nathans. \cite{sn}
 We use the Thomas-Fermi model of the atom to give the
approximate atomic electric field.

\section{Thomas-Fermi model of the atom}

The Thomas-Fermi model of the atom is fully developed in \S70 of \cite{ll}.
Briefly, the atom is treated semi-classically with the electron density as a
function of position determined by phase space considerations. This leads to
a universal function (i.e., it does not depend on atomic number $Z$) for the
self-consistent electric field within the atom, 
\begin{equation}  \label{tfw}
\sqrt{x}\ d^2\chi/dx^2=\chi^{3/2}
\end{equation}
where $\chi$ describes the shielding (assumed spherically symmetric) of the
nuclear point charge, with the boundary conditions that $\chi(0)=1$ and $
\chi(\infty)=0$ (the latter condition determins $\chi^{\prime}(0)$), and the
radius $r$ is related to $x$ as 
\begin{equation}
r=xbZ^{-1/3};\ \ \ b={\frac{1}{2}}({\frac{3}{4}}\pi)^{2/3}=0.885
\end{equation}
(where we are using atomic units so $m_ee^2/\hbar^2=1$). The electric
potential within an atom is given by 
\begin{equation}
\phi(r)={\frac{Ze}{r}}\chi\left({\frac{rZ^{1/3}}{b}}\right)={\frac{Z^{4/3}}{b
}} {\frac{\chi(x)}{x}}.
\end{equation}

We point out that the Thomas-Fermi model does not apply for either very
large or very small $x$; however, the major contribution to the scattering
integral is from $x\approx 1$, and we would expect this model to give
reasonably accurate results.

\section{Interaction of an EDM with the atomic electric field}

We are interested in calculating the spin-dependent neutron scattering
length by the Born approximation for either the $v\times E$ field or a
neutron permanent electric dipole moment (EDM). Consider first the EDM
interaction, 
\begin{equation}
V(r)=-de\vec \sigma\cdot\vec E(r)
\end{equation}
where $d$ is the dipole moment length, $e$ is the magnitude of the electron
charge, $\sigma$ is a Pauli matrix, and $\vec E(r)$ is an electric field.
The scattering amplitude (length) can be determined by use of the Born
approximation, 
\begin{equation}
a=-{\frac{m_n}{2\pi\hbar^2}}\int V(r)e^{i\vec q\cdot \vec r}{\rm d^3}r
\end{equation}
Taking the momentum transfer $\vec q$ along $\hat z$ as the quantization
axis, and using the fact that the electric field is spherically symmetric,
we find 
\begin{equation}
a=\sigma_z{\frac{m}{\hbar^2}}\int_0^\infty deE(r)
e^{iqr\cos\theta}\cos\theta\ \sin\theta\ r^2{\rm d}r{\rm d}\theta
\end{equation}
and the other components are zero because of symmetry. Taking $\vec E
(r)=-\phi^{\prime}(r)\hat r$, and using the Thomas-Fermi wavefunction, the
scattering length can be written as, taking $\beta=bZ^{-1/3}$, 
\begin{equation}
a=-\sigma_z {\frac{m_n}{\hbar^2}}{\beta Ze^2d}\int_0^\infty \left[
-\chi(x)+x\chi^{\prime}(x)\right] e^{ix\beta q\cos\theta}\cos\theta\
\sin\theta {\rm d}x{\rm d}\theta
\end{equation}
which can be rewritten as 
\begin{equation}
a=-\sigma_z i {\frac{m_n}{m_e }}b d Z^{2/3}f(\beta q)
\end{equation}
where $f(\beta q)$ is the imaginary part (the real part is zero) of the
dimensionless integral in the previous equation. The Thomas-Fermi equation
was numerically solved using a Runge-Kutta technique, and the integral
numerically evaluated. The results, as a function of $\beta q$, are shown in
Fig. 1.

For the case of Si, $Z=14$, $\beta=0.367$; a typical $q$ is approximately $
2\pi/2\AA \times 0.5 \AA/a.u.$ giving $\beta q=0.58$, and the dimensionless
integral is about 1. Thus, the difference in the scattering length for the
two spin states (along $\pm\vec q$) is 
\begin{equation}
\Delta a=-2ibdZ^{2/3}m_n/m_e=2\times 10^4 d
\end{equation}
which leads to an spin rotation, on interference with the Si nuclear
scattering amplitude ($a_0=4\times 10^{-13}$ cm) 
\begin{equation}
\Delta \phi={\frac{\Delta a}{a_0}}=4.7\times 10^{16} d/{\rm cm}
\end{equation}
implying that for $d=5\times 10^{-27}$ cm, a Bragg reflection from an Si
crystal would give a rotation of $2\times 10^{-10}$ rad.

\section{$\vec v\times \vec E$ interaction}

Next, consider the $\vec{v}\times \vec{E}$ motional magnetic field
interaction which couples to the neutron magnetic moment, which was first
considered by Schwinger in 1948.\cite{schw}
The possibility of measuring effects from
the motional field has been discussed in regard to non-centrosymmetric
crytals ($\alpha -$quartz) in which case a non-zero average electric field
between scattering planes can exist.\cite{gol,lam} However, as has been
pointed out, there is a $\vec{v}\times \vec{E}$ observable even for
symmetric crystals.\cite{dom}

The hamiltonian for the $\vec{v}\times \vec{E}$ interaction is 
\begin{equation}
V(r)=-\vec{\mu}\cdot \left[ \left( {\frac{\vec{p}}{m_{n}c}}\right) \times 
\vec{E}(\vec{r})\right]/2 .
\end{equation}
For the Born approximation, we use the matrix element 
\begin{equation}
\langle \vec{k}_{2}|V(\vec{r})|\vec{k}_{1}\rangle =\vec{\mu}\cdot {\frac{1}{2
}}\langle \vec{k}_{2}|{\frac{\vec{p}}{m_{n}c}}\times \vec{E}(\vec{r})-\vec{E}
(\vec{r})\times {\frac{\vec{p}}{\vec{m}_{n}c}}|\vec{k}_{1}\rangle 
\end{equation}
where $\vec{k}_{1},\ \vec{k}_{2}$ label the incoming and outgoing neutron
wavefunction. Taking 
\begin{equation}
\vec{v}=\vec{p}/m_{n}={\frac{i\hbar }{m_{n}}}\vec{\nabla}.
\end{equation}
With 
\begin{equation}
\vec{p}|\vec{k}_{1}\rangle =\hbar \vec{k}_{1}|\vec{k}_{1}\rangle ;\ \
\langle \vec{k}_{2}|\vec{p}=\hbar \vec{k}_{2}\langle \vec{k}_{2}|
\end{equation}
gives 
\begin{equation}
\left\langle \vec{k}_{2}|V(\vec{r})|\vec{k}_{1}\right\rangle =-\vec{\mu}
\cdot {\frac{\hbar }{m_{n}c}}{\frac{\vec{k}_{1}+\vec{k}_{2}}{2}}\times \vec{E
}(\vec{r})/2.
\end{equation}
Again, assume that $\vec{q}$ lies along $\hat{z}$; we note that $\vec{q}=
\vec{k}_{2}-\vec{k}_{1}$ is perpendicular to $\vec{k}_{1}+\vec{k}_{2}$
because 
\begin{equation}
(\vec{k}_{2}-\vec{k}_{1})\cdot (\vec{k}_{1}+\vec{k}%
_{2})=k_{2}^{2}-k_{1}^{2}=0
\end{equation}
for elastic scattering. By symmetry, the effective electric field lies along 
$\hat{z}$, and if we assume $\vec{k}_{1}+\vec{k}_{2}$ is along $\hat{y}$ and
$\vec q$
(and has magnitude $2k\cos \theta _{s}$, $2\theta _{s}$ is the scattering
angle), the Born approximation is 
\begin{equation}
a=-\sigma _{x}{\frac{m_{n}}{2\pi \hbar ^{2}}}{\frac {\mu}{2}}k\cos \theta
_{s}\int E(\vec{r})\cos \theta e^{iqr\cos \theta }\sin \theta {\rm d}\theta 
{\rm d}\phi {\rm d}r
\end{equation}
where $\gamma $ is the neutron magnetic moment (-3 Hz/mG). Thus, the
integral is identical to the EDM case, with a different multipliative
constant, and 
\begin{equation}
\Delta a=-2i\pi k\cos \theta _{s}{\frac{\beta Ze}{c}}\mu f(\beta q)
\end{equation}
or a spin rotation about $\hat x$ of 
\begin{equation}
\Delta \phi \approx 4\times 10^{-15}cm/4\times 10^{-13}cm=
10^{-2}\quad rad
\end{equation}
per Bragg reflection from an Si crystal.

\section{Semi-classical model}

If we assume there is no electron cloud around the nucleus,
in Eq. (7) $\chi=1$, and $a_{edm}\propto 1/q$, which is equivalent
to the high-momentum limit.
We can estimate the EDM effect by taking a classical trajectory with impact
parameter ${\rm b}$ relative to the nucleus. The time-integrated-
electric-field-induced phase shift, assuming a spin in the the $\hat z$
direction (propagation in $\hat x$ direction) is given by 
\begin{eqnarray}
\Delta \phi=&-\int_{-\infty}^{\infty} {\frac{edE_z}{\hbar}} {\frac{{\rm d}x}{v
}} =-{\frac{Ze^2d}{\hbar v}}\int_{-\infty}^\infty {\frac{{\rm b}}{({\rm b}
^2+x^2)^{3/2}}}{\rm d}x\\
 =&-{\frac{2Ze^2d}{\hbar {\rm b}v}}=-{\frac{2Ze^2 m_n d
}{\hbar^2 {\rm b} k}}= -2Zd{\frac{m_n}{m_e}}{\frac{1}{k{\rm b}}}
\end{eqnarray}
which is essentially the same as before, in the high momentum limit,
if we take ${\rm b}=a_{nuc}$.

\section{Discussion}

Comparing Eqs. (10) and (19), we see that the motional field spin rotation
is on the order of $10^8$ times larger than that due to an EDM with a
magnitude that would be of interest in an improved experiment.
Unfortunately, the effects cannot be switched on and off as in
the case of the more conventional experiments based on spin precession
in an applied electric field. Although the two scattering
effects are proportional
to $\sigma_x$ and $\sigma_z$ respectively, discrimination between
the effects relies on an absolute determination of the polarization
and scattering axes.  One can also be concerned with the normal nuclear
parity violation, which can combine with a misalignment to produce
effects that mimic T-violation.
This and other issues relevant for a realistic EDM scattering
experiment are similar
to those relating to a study of time reversal violating effects in
slow neutron transmission through polarized matter for which
the issues have been
addressed in some detail; in particular, the constraints on near-perfect
field and polarization alignment, and inability to
discriminate effects due to misalignments,
have been emphasized. \cite{gollam}  The scattering angles
constraints
in a Bragg scattering  EDM experiment are analogous to the 
constraints on the sample 
polarization axis in a neutron transmission experiment as
discussed in \cite{gollam}.  Given the constraints (e.g., scattering
angle and polarization alignment to $10^{-8}$ radian absolute accuracy
which requires $10^{16}$ neutron counts to measure experimentally)
achieving any significant increase in the limit for the neutron EDM
would seem a daunting task.

\begin{figure}
\centerline{\psfig{figure=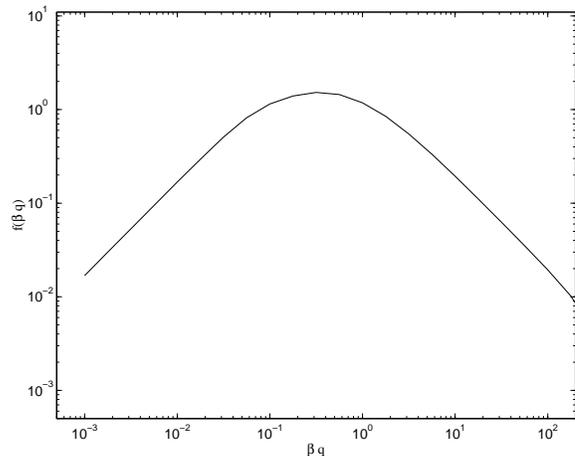,width=3 in}}
\caption{Dimensionless potential integral for the
Born approximation electric field interaction.}
\end{figure}


\begin{references}

\bibitem{sch} C.G. Shull and R. Nathans, Phys. Rev. Lett. {\bf 19}, 384 (1967).

\bibitem{gol} R. Golub and J.M. Pendlebury, Contemp. Phys. {\bf 13}, 519
(1972).

\bibitem{lam}  I.B. Khriplovich and S.K. Lamoreaux, 
\it CP Violation without Strangeness\rm, (Springer-Verlag, Heidelberg,
1997). Sec. 4.3.3.

\bibitem{dom} T. Dombeck, Seminar, Los Alamos National Laboratory,
1999.

\bibitem{schw} Julian Schwinger, Phys. Rev. {73}, 407 (1948).

\bibitem{sn} C.G. Shull, Phys. Rev. Lett. {\bf 10}, 297 (1963).

\bibitem{ll}  L.D. Landau and E.M. Lifshitz, {\it Quantum Mechanics{\rm %
(Pergamon, Oxford, 1977). }}

\bibitem{gollam} S.K. Lamoreaux and R. Golub, Phys. Rev. D {\bf 50},
5632 (1994).

\end{references}
\end{document}